\documentclass[conference, a4paper]{IEEEtran}
\usepackage{tikz,xcolor}
\usepackage[hidelinks, bookmarks=false]{hyperref}
\hypersetup{draft=true}
\usepackage{balance}

\definecolor{lime}{HTML}{A6CE39}
\definecolor{highlight}{RGB}{180, 0, 0}
\DeclareRobustCommand{\orcidicon}{%
	\begin{tikzpicture}
	\draw[lime, fill=lime] (0,0)
	circle [radius=0.16]
	node[white] {{\fontfamily{qag}\selectfont \tiny ID}};
	\draw[white, fill=white] (-0.0625,0.095)
	circle [radius=0.007];
	\end{tikzpicture}
	\hspace{-2mm}
}

\foreach \x in {A, ..., Z}{%
	\expandafter\xdef\csname orcid\x\endcsname{\noexpand\href{https://orcid.org/\csname orcidauthor\x\endcsname}{\noexpand\orcidicon}}
}

\usepackage{cite}
\usepackage{amsmath,amssymb,amsfonts,amsthm}
\usepackage{graphicx}
\usepackage{textcomp}
\usepackage{xcolor}
\usepackage{comment}
\usepackage{float}
\usepackage{wrapfig}
\usepackage{caption}
\usepackage{enumitem}
\def\BibTeX{{\rm B\kern-.05em{\sc i\kern-.025em b}\kern-.08em
    T\kern-.1667em\lower.7ex\hbox{E}\kern-.125emX}}

\usepackage{tabularray}
\usepackage{float}
\usepackage{graphicx}
\usepackage{codehigh}
\usepackage[normalem]{ulem}
\UseTblrLibrary{booktabs}
\UseTblrLibrary{siunitx}

\NewTableCommand{\tinytableDefineColor}[3]{\definecolor{#1}{#2}{#3}}

\newtheorem{Proposition}{Proposition}
\newtheorem{Definition}{Definition}
\newtheorem{Example}{Example}
\newtheorem{Theorem}{Theorem}

\newcommand{\routid}[1]{\mathbf{R}_{\mathrm{Id}}(#1)}
\newcommand{\routuni}[1]{\mathbf{R}_{\mathrm{Uni}}(#1)}
\newcommand{\routcut}[1]{\mathbf{R}_{\mathrm{Cut}}(#1)}
\newcommand{\hypid}[1]
{\mathcal{H}_{\mathrm{Id}}(#1)}
\newcommand{\hypuni}[1]{\mathcal{H}_{\mathrm{Uni}}(#1)}
\newcommand{\hypcut}[1]{\mathcal{H}_{\mathrm{Cut}}(#1)}

\usepackage{subcaption}
\usepackage{booktabs}
\usepackage{stfloats}
\usepackage{algorithm}
\usepackage{algpseudocode}
\usepackage{tabto}

\begin{document}

\title{Methods for Path Set Attribute Calculation in Network Systems}

\makeatletter
\newcommand{\linebreakand}{%
  \end{@IEEEauthorhalign}
  \hfill\mbox{}\par
  \mbox{}\hfill\begin{@IEEEauthorhalign}
}
\makeatother

\author{
\IEEEauthorblockN{Giovanni Fiaschi\orcidA{}}
\IEEEauthorblockA{\textit{Radio New Concept and Algorithms} \\
\textit{Ericsson AB}\\
Stockholm, Sweden\\
giovanni.fiaschi@ericsson.com}
\and
\IEEEauthorblockN{Carlo Vitucci\orcidB{}}
\IEEEauthorblockA{\textit{Technology Management} \\
\textit{Ericsson AB}\\
Stockholm, Sweden \\
carlo.vitucci@ericsson.com}
\and
\IEEEauthorblockN{Thomas Westerbäck\orcidC{}}
\IEEEauthorblockA{\textit{Division of Mathematics and Physics} \\
\textit{Mälardalen University}\\
Västerås, Sweden \\
thomas.westerback@mdu.se}
\linebreakand
\IEEEauthorblockN{Daniel Sundmark\orcidD{}}
\IEEEauthorblockA{\textit{Computer Science and Software Enigineering} \\
\textit{Mälardalen University}\\
Västerås, Sweden \\
daniel.sundmark@mdu.se}
\and
\IEEEauthorblockN{Thomas Nolte\orcidE{}}
\IEEEauthorblockA{\textit{Division of Networked and Embedded System} \\
\textit{Mälardalen University}\\
Västerås, Sweden \\
thomas.nolte@mdu.se}
}

\maketitle

\begin{abstract}
In graph theory and its applications to networking, such as telecommunications or transportation, path-finding is a central problem. While single-path algorithms are well established, methods for handling sets of multiple paths are less developed. A companion paper introduced a formal model for defining attributes over sets of paths based on their structural properties; this paper addresses that model's practical implementation. We present an optimized algorithm for computing cut sets of a path set—a nontrivial task that can be infeasible without efficient methods—and validate its performance via systematic benchmarks on network simulations of varying complexity. Additionally, we introduce a vectorized computational framework that expresses property calculations as matrix operations, enabling concise implementations in array-oriented languages. Together, these contributions establish practical foundations for the companion model, demonstrating that its implementation is both feasible and characterized by predictable, acceptable execution times.
\end{abstract}


\section{Introduction}
\label{Sec:Introduction}

Routing is about finding paths over graphs~\cite{Medhi2017}. Path search is widely used in telecommunications and transportation networks, and can also help find connections between graph vertices when graphs represent other domains~\cite{Zhou2023, Cheung2018}.
To find a path, it is assumed that the graph is weighted, that is, each edge of the graph has an associated real value, and that a corresponding value can be calculated for a path from the edges it crosses~\cite{Mathew2023}. This value is referred to as 'length' or 'cost' in the shortest path algorithms, the most widely used algorithms in the path search~\cite{Dijkstra1959, Bellman1958, Ford1956}. Other properties may be interesting when connecting two vertices in a graph. Some of these properties, such as capacity or fault probability, may be improved by using multiple paths. For example, a set of several paths reduces the probability that a fault will disrupt the connectivity between the two vertices.

To the best of our knowledge, the characterization of path properties of multipath sets has received too little attention in the literature. Existing approaches focus always on cost and diversity~\cite{Yen1971, Bhandari1999, Liu2018, Chondrogiannis2015, Fiaschi2020}, ignoring the fact that a path set is typically designed to enhance a specific functional property, such as reliability or throughput. This instead would require metrics and methods explicitly tailored to that objective.\\

Fiaschi et al.~\cite{Fiaschi2025} proposed a systematic way to calculate the properties of the path set, using a serial composition operator, a parallel composition operator, and, if needed, a transformation of the representation of the path set. This was applied to five heterogeneous property examples for illustrative purposes, namely delay, cost, capacity, unavailability, and fault probability, and showed that:
\begin{itemize}
    \item The property calculation is specific to the property, that is, different properties require different calculations.
    \item The calculation always requires a serial operator, a parallel operator and a path set representation, possibly transformed, that can be conveniently expressed as a matrix.
\end{itemize}

\cite{Fiaschi2025} introduced a mathematical model to represent the intrinsic contributions of the path set:
\begin{itemize}
    \item an operation representing the contribution of the serialization of the edges in a path ($OP_s$),
    \item an operation representing the contribution of the parallelization of the paths in a set of disjoint paths ($OP_p$),
    \item a transformation to combine the two operations in an arbitrary path set ($T$).
\end{itemize}

Equation~\ref{equ:MathFramework} summarizes these three elements of the model:
\begin{equation}
    \label{equ:MathFramework}
    \Psi(P) = OP_s \circ OP_p \circ T
\end{equation}
Table~\ref{tab:transformationsummary} shows the serial and parallel operations and the path set transformation for delay, cost, capacity and fault probability. The notation used for the transformation will be clarified in section~\ref{sec:vertex-weighted-hypergraphs}.
\\

This paper introduces practical methods to implement the model of Equation~\ref{equ:MathFramework} by using tools that are commonly used in modern programming languages, such as Hadamard operations, and by describing algorithms to derive the intermediate representations, such as the cut matrix.
\\

Section~\ref{sec:previous} recalls the framework and concepts from~\cite{Fiaschi2025}, for completeness.
Section~\ref{sec:transformations} discusses algorithms to compute minimal cuts of a path set, as this is the least trivial of the path set transformations.
Section~\ref{sec:benchmark} provides execution times of implementations of the above algorithms in notable examples, to demonstrate their practical tractability.
Section~\ref{sec:hadamard} introduces the Hadamard product and power, as well as broadcasting, which is common in modern programming languages and useful for the implementation of the methods in \cite{Fiaschi2025}.
In Section~\ref{sec:equivalence} we prove the correlation between the \( r \)-incidence matrices and the Hadamard operations. Finally, Section~\ref{sec:implementation} shows the overall implementation of the path set characterization.

\begin{table}[!t]
\centering
\small
\resizebox{\columnwidth}{!}{%
\begin{tabular}{||l|ccc||}
 \hline
 \hline
 Example & \rule{0pt}{2.5ex}$OP_s$ & $OP_p$ & T \\ [0.5ex]
 \hline
 Delay & \rule{0pt}{2.5ex} $+$ & $max$ & $R(0, w, \routid{P})$ \\
\hline
 Administrative & \rule{0pt}{2.5ex} $+$ & $+$ & $R(0, w, \routuni{P})$ \\
 cost&  &  &  \\
 \hline
 Combined & \rule{0pt}{2.5ex} $min$ & $+$ & $R(0, w, \routcut{P})$ \\
 capacity&  &  &  \\
 \hline
 Fault& \rule{0pt}{2.5ex} $+$ & $\times$ & $R(1, w, \routcut{P})$ \\
 probability&  &  &  \\
 \hline
 \hline
\end{tabular}
}
\caption{Operations and transformations, according to~\cite{Fiaschi2025}}
\label{tab:transformationsummary}
\end{table}

\section{Definitions} \label{sec:previous}

This section recalls concepts and definitions used later on in the paper. The following concepts are not further detailed: \emph{graph} $\mathcal{G} = (V_g, E_g)$ ($V_g$ vertices and $E_g$ edges), \emph{path} $p \subseteq E_g$ (set of contiguous edges), \emph{routing table} $R = [r_{ij}]$ of a set of $k$ paths $P=\{p_1, \dots, p_k \}$ ($r_{ij}=1$ if edge $e_i \in p_j$, $r_{ij}=0$ otherwise), and \emph{edge property vector} $\mathbf{w} \in \mathbb{R}^{|E_g|}$ (encoding the edge properties of a graph). Table \ref{tab:PathSet_Transformation_tables}.A shows an edge property vector $\mathbf{w} = (w_1,\ldots,w_8)$ for the graph in Figure~\ref{fig:3paths}.
For more complete definitions refer to Fiaschi et al.~\cite{Fiaschi2025}.

\subsection{Hypergraphs}

Hypergraphs were introduced by C.~Berge in the 1960s~\cite{berge1984hypergraphs} as a generalization of graphs where edges join any number of vertices.

A (finite) \textit{hypergraph} is the pair $\mathcal{H} = (V_h,E_h)$ where $V_h$ is a finite set of elements called \textit{vertices}, $E_h$ is a finite set of elements called \textit{edges}, and $e \subseteq V_h$ for each element $e \in E_h$.
Hypergraphs can be represented using incidence matrices.
The \textit{incidence matrix} $B=(b_{ij})_{m \times n}$ of a hypergraph $\mathcal{H} = (V_h,E_h)$ with $V_h=\{v_1,\ldots,v_m\}$ and $E_h=\{e_1,\ldots,e_n\}$ is defined by\\
\[
b_{ij} = \biggl\{
\begin{array}{ll}
1 & \hbox{if $v_i \in e_j$,}\\
0 & \hbox{otherwise.}
\end{array}
\]

\subsection{Path Sets}
\label{sec:PathSets}

\begin{figure}[!b]
\centering
    \includegraphics[width=1\linewidth, keepaspectratio]{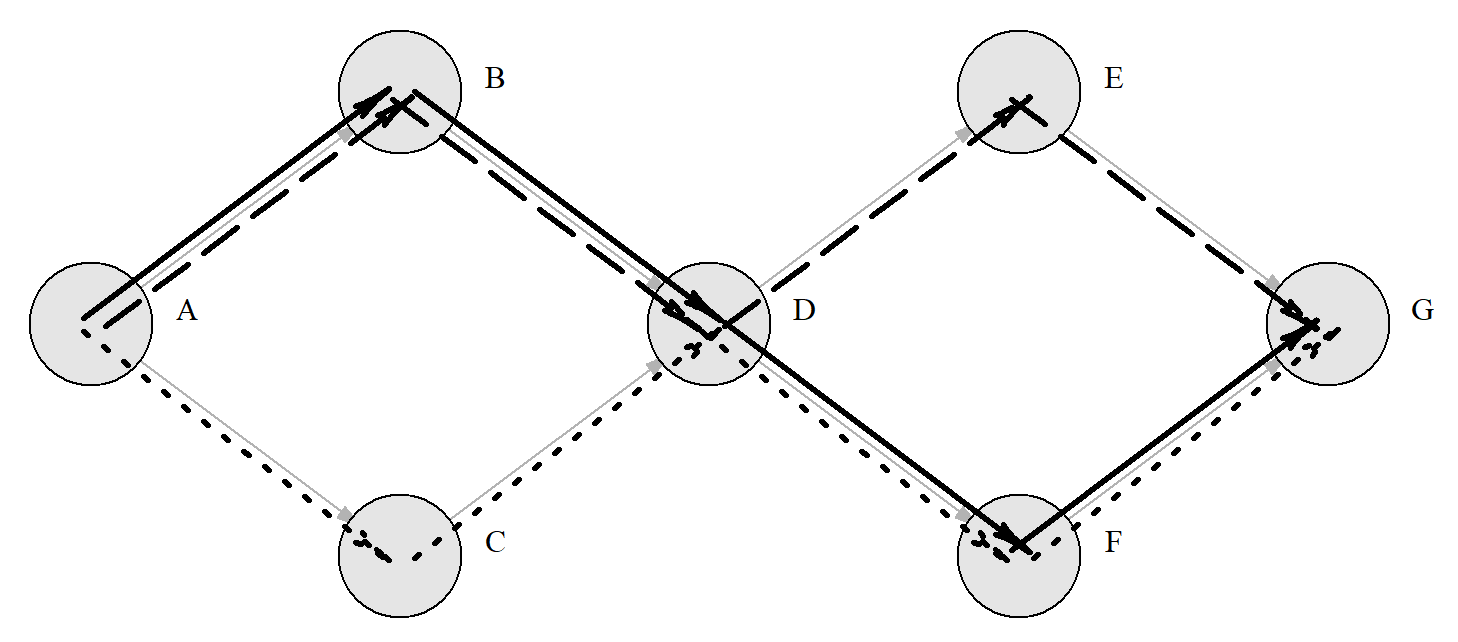}
  \caption{A graph with three paths}%
  \label{fig:3paths}
\end{figure}

A set of directed paths can be described by a hypergraph, whose vertices correspond to the edges of the graph on which the paths are routed and whose edges correspond to the paths in the set.

In the following, the notation $\hypid{P}$ indicates the hypergraph describing path set $P$ and $\routid{P}$ its incidence matrix.

\begin{Example}
\label{ex:incidencematrix_id}
Let $\mathcal{G}=(V_g, E_g)$ be the graph of Figure \ref{fig:3paths}, so that
\NumTabs{8}
\begin{itemize}
    \item $V_g=\{A, B, C, D, E, F, G\}$
    \item $E_g=\{A \rightarrow B, A \rightarrow C, B \rightarrow D, C \rightarrow D,$ \\
    \tab \tab $D \rightarrow E, D \rightarrow F, E \rightarrow G, F \rightarrow G\}$
\end{itemize}
Let $P=\{p_1, p_2, p_3\}$ be the set of the three paths in the same figure, respectively dashed, solid and dotted. Let $\hypid{P}=(V_h, E_h)$ be the hypergraph representing $P$ on $\mathcal{G}$, so that $V_h=E_g$ and $E_h=P$. Then, the incidence matrix $B=\routid{P}$ of $\hypid{P}$ is equal to the routing table of $P$ in $\mathcal{G}$ and is given in Table \ref{tab:PathSet_Transformation_tables}.B.
\end{Example}

\subsection{Union}
\label{sec:Union}

A very simple transformation of a path set $P$ is the union of all the edges pertaining to its member paths. The union of a path set is a single set of edges, therefore it can be represented by a hypergraph with a single edge.

The notations $\hypuni{P}$ and $\routuni{P}$ will indicate the hypergraph of the union of $P$ and its incidence matrix respectively.

The union can be easily calculated from the routing matrix by taking the row-wise logical OR of its entries.

\begin{Example}
\label{ex:incidencematrix_uni}
Let $\mathcal{G}=(V_g, E_g)$ and $P$ be respectively the graph and path set as in Example \ref{ex:incidencematrix_id}.
Let $\hypuni{P}=(V_h, E_h)$ be the hypergraph representing the union of $P$ on $\mathcal{G}$, so that $V_h=E_g$ and $E_h$ is the set whose single element is the union of all paths in $P$. Then, the incidence matrix $B=\routuni{P}$ of $\hypuni{P}$ has a single column and is given in Table \ref{tab:PathSet_Transformation_tables}.C. It has ones in all positions, as at least one path crosses each edge of the graph in this example.
\end{Example}

\begin{table*}[!t]
\centering
\resizebox{\linewidth}{!}{%
\begin{tabular}{||l || c || c c c || c || c c c c c c c c c c c c || }
\hline
\hline
& A & \multicolumn{3}{c ||}{B} & C & \multicolumn{12}{c ||}{D} \\
\hline
& $\mathbf{w}$ & \multicolumn{3}{c ||}{$\routid{P}$} & $\routuni{P}$ & \multicolumn{12}{c ||}{$\routcut{P}$}\\
\hline
    Edge & Weight & $p_1$ & $p_2$ & $p_3$ & Union & $C_1$ & $C_2$ & $C_3$ & $C_4$ & $C_5$ & $C_6$ & $C_7$ & $C_8$ & $C_9$ & $C_{10}$ & $C_{11}$ & $C_{12}$ \\
\hline
    A→B & $w_1$ & 1 & 1 & 0 & 1 & 1 &   1 &   1 &   1 &   0 &   0 &   0 &   0 &   0 &   0 &   0 &   0 \\
    A→C & $w_2$ & 0 & 0 & 1 & 1 & 1 &   0 &   0 &   0 &   1 &   0 &   0 &   0 &   0 &   0 &   0 &   0 \\
    B→D & $w_3$ & 1 & 1 & 0 & 1 & 0 &   0 &   0 &   0 &   1 &   1 &   1 &   1 &   0 &   0 &   0 &   0 \\
    C→D & $w_4$ & 0 & 0 & 1 & 1 & 0 &   1 &   0 &   0 &   0 &   1 &   0 &   0 &   0 &   0 &   0 &   0 \\
    D→E & $w_5$ & 1 & 0 & 0 & 1 & 0 &   0 &   0 &   0 &   0 &   0 &   0 &   0 &   1 &   1 &   0 &   0 \\
    D→F & $w_6$ & 0 & 1 & 1 & 1 & 0 &   0 &   1 &   0 &   0 &   0 &   1 &   0 &   1 &   0 &   1 &   0 \\
    E→G & $w_7$ & 1 & 0 & 0 & 1 & 0 &   0 &   0 &   0 &   0 &   0 &   0 &   0 &   0 &   0 &   1 &   1 \\
    F→G & $w_8$ & 0 & 1 & 1 & 1 & 0 &   0 &   0 &   1 &   0 &   0 &   0 &   1 &   0 &   1 &   0 &   1 \\
    \hline
    \hline
    \end{tabular}%
    }
    \caption{Path set transformation tables}
    \label{tab:PathSet_Transformation_tables}
\end{table*}

\subsection{Cuts}
\label{sec:Cuts}

A cut is a well-known concept in graph theory: given a partition of the vertices into two disjoint subsets, a graph cut is the set of all edges that have one end vertex in each partition of the cut~\cite{Diestel2000}.

Given a path set connecting two vertices $S$ and $D$, \cite{Fiaschi2025} introduced the expression \textit{cut of the path set} for a minimal edge set that, if removed, will make $S$ and $D$ unreachable within the path set. The cut is minimal in the sense that all the edges in the cut must be removed from the path set to make the vertices unreachable.

For example, in the paths of Figure~\ref{fig:3paths}, the edge set $\{B \rightarrow D, F \rightarrow G\}$ is:
\begin{itemize}
    \item a minimal cut of the path set $P = \{p_1, p_2, p_3\}$, as removing all its edges would make all the paths in $P$ unusable and removing only one of its edges would leave at least one path in $P$ connecting $A$ to $G$
    \item not a minimal cut of graph $\mathcal{G}$ because with a path $p_4 = \{A \rightarrow C, C \rightarrow D, D \rightarrow E, E \rightarrow G\}$, which is not in $P$, it would still be possible to reach $G$ from $A$ in the graph
\end{itemize}

The notations $\hypcut{P}$ and $\routcut{P}$ will indicate the hypergraph of the cuts of $P$ and its incidence matrix respectively.

\begin{Example}
\label{ex:incidencematrix_cut}
Let $\mathcal{G}=(V_g, E_g)$ and $P$ be respectively the graph and path set as in Example \ref{ex:incidencematrix_id}.
Let $\hypcut{P}=(V_h, E_h)$ be the hypergraph representing the set of minimal cuts of $P$ on $\mathcal{G}$, so that $V_h=E_g$ and $E_h$ is the set of all minimal cuts of $P$ in $\mathcal{G}$. Then, the incidence matrix $B=\routcut{P}$ of $\hypcut{P}$ is given in Table \ref{tab:PathSet_Transformation_tables}.D.
\end{Example}

\subsection{Vertex-Weighted Hypergraphs}
\label{sec:vertex-weighted-hypergraphs}
Vertex-weighted hypergraphs are a generalization of hypergraphs. For vertex-weighted hypergraphs, \cite{Fiaschi2025} also introduces \( r \)-incidence matrices, which generalize the concept of incidence matrices for hypergraphs.

A \textit{vertex-weighted hypergraph} is a tuple $\mathcal{H}_w = (V_h, E_h, \mathbf{w})$ where $\mathcal{H}=(V_h,E_h)$ is a hypergraph and $\mathbf{w} = [w_1, \ldots w_{|V_h|}]$ is a weight vector in $\mathbb{R}^{|V_h|}$ which assign a real number $w_i$ to each vertex $v_i \in V_h$. $\mathcal{H}$ is said to be the \emph{underlying hypergraph} of $\mathcal{H}_w$.

If a hypergraph represents a set of sets of edges in a graph $\mathcal{G}$, and $\mathcal{G}$ has an associated edge property vector $\mathbf{w}$, then a vertex-weighted hypergraph can be easily constructed by associating $\mathbf{w}$ to the vertices of the hypergraph, since they correspond to the edges of $\mathcal{G}$.
\hfill\break

Given a vertex-weighted hypergraph $\mathcal{H}_w = (\mathcal{H}, \mathbf{w}) = (V_h,E_h,\mathbf{w})$ with $V_h = \{v_1, \ldots, v_m\}$ and $E_h = \{e_1, \ldots, e_n\}$, for any real number $r \in \mathbb{R}$, let define the \textit{r-incidence matrix} $B_r = (b_{ij})_{m \times n}$ as:\\
\[
b_{ij} = \biggl\{
\begin{array}{ll}
w_i & \hbox{if $v_i \in e_j$,}\\
r & \hbox{otherwise.}
\end{array}
\]

The notation $B_r = R(r, \mathbf{w}, B)$ denotes the \(r\)-incidence matrix of a vertex-weighted hypergraph $\mathcal{H}_w = (\mathcal{H}, \mathbf{w})$, where $B$ represents the incidence matrix of the underlying hypergraph $\mathcal{H}$.

\begin{Example} \label{ex:incidencematrix_weighted}
Let $\mathcal{H}_c = (\hypcut{P}, \mathbf{c})$ be the vertex-weighted hypergraph, whose underlying hypergraph is $\hypcut{P}$ as given in Example \ref{ex:incidencematrix_cut}, and
$\mathbf{c}$ the capacity vector given in Table \ref{tab:r-incidence}.A. Then the 0-incidence matrix $B_0=R(0, \mathbf{c}, \routcut{P})$ of $\mathcal{H}_c$ is given in Table \ref{tab:r-incidence}.B.
\end{Example}

\begin{Example} \label{ex:incidencematrix_probability_weighted}
Let $\mathcal{H}_p = (\hypcut{P}, \mathbf{p})$ be the vertex-weighted hypergraph, whose underlying hypergraph is $\hypcut{P}$ as given in Example \ref{ex:incidencematrix_cut}, and
$\mathbf{p}$ the fault probability vector given in Table \ref{tab:r-incidence}.A. Then the 1-incidence matrix $B_1=R(1, \mathbf{p}, \routcut{P})$ of $\mathcal{H}_p$ is given in Table \ref{tab:r-incidence}.C.
\end{Example}

\begin{table*}[!t]
\centering
\setlength{\tabcolsep}{2pt}
\resizebox{\linewidth}{!}{%
\begin{tabular}{|| l || c c || c c c c c c c c c c c c || c c c c c c c c c c c c || }
\hline
\hline
& \multicolumn{2}{c ||}{A} & \multicolumn{12}{c ||}{B} & \multicolumn{12}{c ||}{C} \\
\hline
& Cap. & Prob. & \multicolumn{12}{c ||}{$R(0, c, \routcut{P})$} & \multicolumn{12}{c ||}{$R(1, p, \routcut{P})$} \\
\hline
    Edge & $\mathbf{c}$ & $\mathbf{p}$ & C1 & C2 & C3 & C4 & C5 & C6 & C7 & C8 & C9 & C10 & C11 & C12 & C1 & C2 & C3 & C4 & C5 & C6 & C7 & C8 & C9 & C10 & C11 & C12 \\
\hline
    A→B &
    25 & 0.0050 &
    25 & 25 & 25 & 25 & 0 & 0 & 0 & 0 & 0 & 0 & 0 & 0 &
    0.0050 & 0.0050 & 0.0050 & 0.0050 & 1 & 1 & 1 & 1 & 1 & 1 & 1 & 1 \\
    A→C &
    10 & 0.0075 &
    10 & 0 & 0 & 0 & 10 & 0 & 0 & 0 & 0 & 0 & 0 & 0 &
    0.0075 & 1 & 1 & 1 & 0.0075 & 1 & 1 & 1 & 1 & 1 & 1 & 1 \\
    B→D &
    25 & 0.0070 &
    0 & 0 & 0 & 0 & 25 & 25 & 25 & 25 & 0 & 0 & 0 & 0 &
    1 & 1 & 1 & 1 & 0.0070 & 0.0070 & 0.0070 & 0.0070 & 1 & 1 & 1 & 1 \\
    C→D &
    25 & 0.0040 &
    0 & 25 & 0 & 0 & 0 & 25 & 0 & 0 & 0 & 0 & 0 & 0 &
    1 & 0.0040 & 1 & 1 & 1 & 0.0040 & 1 & 1 & 1 & 1 & 1 & 1 \\
    D→E &
    10 & 0.0115 &
    0 & 0 & 0 & 0 & 0 & 0 & 0 & 0 & 10 & 10 & 0 & 0 &
    1 & 1 & 1 & 1 & 1 & 1 & 1 & 1 & 0.0115 & 0.0115 & 1 & 1 \\
    D→F &
    25 & 0.0045 &
    0 & 0 & 25 & 0 & 0 & 0 & 25 & 0 & 25 & 0 & 25 & 0 &
    1 & 1 & 0.0045 & 1 & 1 & 1 & 0.0045 & 1 & 0.0045 & 1 & 0.0045 & 1 \\
    E→G &
    10 & 0.0105 &
    0 & 0 & 0 & 0 & 0 & 0 & 0 & 0 & 0 & 0 & 10 & 10 &
    1 & 1 & 1 & 1 & 1 & 1 & 1 & 1 & 1 & 1 & 0.0105 & 0.0105 \\
    F→G &
    100 & 0.0065 &
    0 & 0 & 0 & 100 & 0 & 0 & 0 & 100 & 0 & 100 & 0 & 100 &
    1 & 1 & 1 & 0.0065 & 1 & 1 & 1 & 0.0065 & 1 & 0.0065 & 1 & 0.0065 \\
    \hline
    \hline
    \end{tabular}%
    }
     \caption{\textit{r}-incidence matrices}
    \label{tab:r-incidence}
\end{table*}

\section{Calculation of the Minimal Cuts of a Path Set}
\label{sec:transformations}

A trivial way to calculate the cuts of a path set is to take the $n$-ary Cartesian product of the paths in the set, that is all the tuples with one edge from each path, then interpret the resulting tuples as sets and perform the following simplifications:
\begin{itemize}
    \item eliminate duplicate elements in each cut (as a cut is a set of edges, not a multiset),
    \item eliminate duplicate cuts in the cut set,
    \item eliminate cuts that are supersets of other cuts in the cut set (eliminate non-minimal cuts).
\end{itemize}

For sets of fully disjointed paths, no simplification is needed, and the computational complexity, both in time and memory, is $\mathcal{O}(\mathit{ProdL})$, where
\hfill\break

$
\begin{array}{ll}
\mathit{ProdL} & = \prod_{i=1}^{k}{l_i}\\
l_i & =\text{ number of edges in path } i\\
k & = \text{ number of paths in the path set}
\end{array}
$
\hfill\break

Considering then the simplifications, the superset elimination is the most complex. Assuming efficient representation of cuts as sets of edges allowing constant time test for supersets, if all the cuts must be pairwise compared, the complexity is in the order of the square of the initial number of cuts, therefore a worst-case upper bound on the complexity of the cut set calculation is $\mathcal{O}(\mathit{ProdL}^2)$.
\hfill\break
\hfill\break

A recursive approach may be more clever, performing the checks during the search, therefore pruning the space in case of partially overlapping paths and anyway avoiding the need of the last check, so improving the average complexity in general and dropping the worst-case back to $\mathcal{O}(\mathit{ProdL})$.
\hfill\break

A sketch of the algorithm is given in the pseudocode of Algorithm~\ref{algor:cuts}. The procedure \textsc{cuts\_helper} is the recursive procedure that keeps track of edges to be skipped in extra parameters $v$ and $\mathcal{C}$ (explained below). This procedure is called in the main procedure \textsc{cuts} with $P$ containing the full path set to be evaluated, $v = \varnothing$ and $\mathcal{C} = \varnothing$, that is, at the beginning, no edge must be skipped.
\hfill\break

The recursive procedure \textsc{cuts\_helper} receives four arguments, $P_0$ (the initial path set, passed unchanged through recursions), $P$ (the set of paths recursively reduced), $v$ (a set of edges in the graph to be skipped because already considered in previous calls), and $\mathcal{C}$ (a set of sets of paths, each set cut by one edge in previous recursions), and works as follows:
\begin{itemize}
    \item Pick one path $p$ from $P$ (line 4) and create as many cut sets as edges in $p$ (excluding edges to be skipped), inserting the edges one per cut (loop at line 8).
    \item For each cut, add to the already inserted edge $e$ the edges that disrupt all the paths in $P$ not crossing $e$. This is done by calling recursively the function \textsc{cuts\_helper} only on a subset of $P$. Therefore, \textsc{cuts\_helper} (see line 15) is called with second parameter equal to all the paths $p_1$ in $P$ that do not contain $e$.
    \item Keep track of the edges already considered in previous recursions in a set $v$.
    The edges in $v$ will be skipped in subsequent recursive calls.
    The loop at line 8 is over the set $d$ defined in line 5, which contains all the edges of $p$ that are not in $v$. Instead, in the recursive call of \textsc{cuts\_helper} at line 15, the edges to be skipped will be $u$, defined at lines 6 and 13 as $v$ plus all the edges of $p$ visited in previous iterations of loop at line 8.
    \item Keep track of the sets of paths that are already cut by previously visited edges in the set of sets of paths $\mathcal{C}$. The set $c_e$ of paths from the entire initial path set $P_0$ that are cut by the edge $e$ under examination in the current iteration is calculated in line 9. If, by adding $e$, edges added to the cut in previous recursions become redundant (test in line 10), $e$ is skipped, as it would otherwise generate a cut that is not minimal.
\end{itemize}

\begin{algorithm}
\caption{Cut Set Calculation}
\label{algor:cuts}
\begin{algorithmic}[1]
    \Procedure{cuts\_helper}{$P_0, P, v, \mathcal{C}$}
      \If {$P=\varnothing$}
        \Return $\varnothing$
      \EndIf
      \State $p \gets$ first$(P)$
      \State $d \gets p \setminus v$ \Comment set difference between $p$ and $v$
      \State $u \gets v$
      \State $C \gets \varnothing$
      \ForAll{$e \in d$}
        \State $c_e \gets \{p_0 \in P_0 \mid e \in p_0\}$
        \Comment paths in $P_0$ crossing $e$
        \If {$\exists c_x \in \mathcal{C} \mid c_x \subseteq \left( \bigcup_{c \in \mathcal{C} \setminus \{c_x\}}c \right) \cup c_e$} \\
          \quad \quad \quad \quad \textbf{continue} \Comment skip edge $e$ due to non-minimal cut
        \EndIf
        \State $u \gets u \cup \{e\}$
        \State $C_1 \gets $ \\ \quad \quad \quad \quad $\Call{cuts\_helper}{
        P_0, P \setminus c_e,
        u, \mathcal{C} \cup \{c_e\}}$
        \If {$C_1 = \varnothing$}
          \Return $\varnothing$
        \EndIf
        \ForAll{$c \in C_1$}
          \State $c \gets c \cup \{e\}$
        \EndFor
        \State $C \gets C \cup C_1$
      \EndFor
      \State \Return $C$
    \EndProcedure
\\
    \Procedure{cuts}{$P$}
      \State \Return $\Call{cuts\_helper}{
      P, P, \varnothing, \varnothing}$
    \EndProcedure
\end{algorithmic}
\end{algorithm}

\section{Cut Calculation Benchmarking}
\label{sec:benchmark}

\begin{table*}
\centering
\begin{tabular*}{\textwidth}{@{\extracolsep{\fill}} lrllrrrr}
\toprule
Path endpoints & $k$ & $\lambda$ & Path lengths & ProdL & CutNo & \textsc{Sys} median (ms) & \textsc{Red} median (ms)\\
\midrule
v91→v92 (red dots) & 3 & low & 9 10 9 & 810 & 225 & 13.95 & 12.52\\
v91→v92 & 4 & low & 9 10 9 10 & 8100 & 897 & 108.66 & 68.02\\
v91→v92 & 5 & low & 9 10 9 10 9 & 72900 & 1887 & 1095.85 & 256.61\\
v91→v92 & 3 & high & 9 10 14 & 1260 & 1050 & 42.03 & 50.42\\
v91→v92 & 4 & high & 9 10 14 16 & 20160 & 15610 & \textbf{9379.72} & 1024.67\\
v91→v92 & 5 & high & 9 10 14 16 14 & 282240 & 110974 & \textbf{\underline{671967.20}} & \textbf{14517.65}\\
v100→v38 (green dots) & 3 & low & 3 3 3 & 27 & 9 & 0.40 & 0.57\\
v100→v38 & 4 & low & 3 3 3 5 & 135 & 45 & 1.67 & 2.53\\
v100→v38 & 5 & low & 3 3 3 5 5 & 675 & 135 & 7.54 & 9.74\\
v100→v38 & 3 & high & 3 3 5 & 45 & 45 & 0.98 & 2.02\\
v100→v38 & 4 & high & 3 3 5 32 & 1440 & 1125 & 44.01 & 65.78\\
v100→v38 & 5 & high & 3 3 5 32 19 & 27360 & 12747 & \textbf{7628.56} & 1033.89\\
\bottomrule
\end{tabular*}
\caption{Test cases}
\label{tab:test_cases}
\end{table*}

\begin{table*}[!b]
\centering
\begin{talltblr}[
entry=none,label=none,
note{}={Est.: Coefficient; S.E.: Standard Error; t: t-statistic; p: p-value.},
note{ }={Stars indicate statistical significance levels: + p \num{< 0.1}, * p \num{< 0.05}, ** p \num{< 0.01}, *** p \num{< 0.001}},
]
{
colspec={Q[]Q[]Q[]Q[]Q[]Q[]Q[]Q[]Q[]},
hline{2}={3-4,7-9}{solid, black, 0.03em},
hline{2}={2,6}{solid, black, 0.03em, l=-0.5},
hline{2}={5}{solid, black, 0.03em, r=-0.5},
hline{3}={1-9}{solid, black, 0.05em},
hline{6}={1-9}{solid, black, 0.05em},
hline{1}={1-9}{solid, black, 0.08em},
hline{9}={1-9}{solid, black, 0.08em},
column{3-5,7-9}={}{halign=c},
cell{1}{1}={}{halign=c},
cell{1}{2}={c=4}{halign=c},
cell{1}{6}={c=4}{halign=c},
cell{2-8}{1}={}{halign=l},
cell{2-8}{2}={}{halign=c},
cell{2-8}{6}={}{halign=c},
}
& Sys &  &  &  & Red &  &  &  \\
& Est. & S.E. & t & p & Est. & S.E. & t & p \\
(Intercept) & \num{-5.435}*** & \num{0.616} & \num{-8.828} & \num{<0.001} & \num{-3.429}*** & \num{0.142} & \num{-24.131} & \num{<0.001} \\
log(ProdL) & \num{0.458}* & \num{0.194} & \num{2.360} & \num{0.046} & \num{0.230}** & \num{0.049} & \num{4.687} & \num{0.001} \\
log(CutNo) & \num{0.958}** & \num{0.215} & \num{4.455} & \num{0.002} & \num{0.847}*** & \num{0.052} & \num{16.242} & \num{<0.001} \\
Num.Obs. & \num{11} &  &  &  & \num{12} &  &  &  \\
R2 & \num{0.970} &  &  &  & \num{0.997} &  &  &  \\
R2 Adj. & \num{0.963} &  &  &  & \num{0.997} &  &  &  \\
\end{talltblr}
\caption{Linear models}
\label{tab:linear_models}
\end{table*}

In the previous section, two algorithms were presented to calculate minimal cuts in a path set:
\begin{itemize}
    \item one systematically generating all the edge combinations and then removing duplicate or non-minimal cuts, referred as \textsc{Sys} in this section,
    \item one using a recursively reduced search, Algorithm \ref{algor:cuts}, referred here as \textsc{Red}.
\end{itemize}

The theoretical complexity estimates, $\mathcal{O}(\mathit{ProdL}^2)$ for \textsc{Sys} and $\mathcal{O}(\mathit{ProdL})$ for \textsc{Red}, underwent empirical verification using actual algorithm runs. Although the sample size remains relatively small, the results provide meaningful insights into the scaling behavior.

The experimental setup utilized a random network graph (96 vertices and 322 edges, Figure \ref{fig:net96}) generated using the Waxman method \cite{Waxman1988}. Two pairs of endpoint vertices served as the basis for 12 test cases (Table~\ref{tab:test_cases}), created by varying the number of paths ($k=3, 4, 5$) and a disjointedness parameter (low/high value of $\lambda$), using the algorithm described in \cite{Fiaschi2020}.

Validation of the complexity estimates relied on recording execution times and confirming their alignment with $\mathcal{O}(\mathit{ProdL}^2)$ and $\mathcal{O}(\mathit{ProdL})$ through log-log linear regression. The analysis reveals that the final number of cuts ($\mathit{CutNo}$) acts as a more precise predictor of execution time than $\mathit{ProdL}$. However, because $\mathit{CutNo}$ is unknown until the algorithm terminates, $\mathit{ProdL}$ must be used for an \textit{a priori} complexity statement. Since $\mathit{CutNo} \leq \mathit{ProdL}$ by construction, substituting $\mathit{ProdL}$ for $\mathit{CutNo}$ yields a conservative upper-bound estimate ($\mathcal{O}(\mathit{ProdL} \cdot \mathit{CutNo}) \subseteq \mathcal{O}(\mathit{ProdL}^{2})$).
Table~\ref{tab:test_cases} explicitly reports $\mathit{ProdL}$, $\mathit{CutNo}$, and the measured execution times for both algorithms across all test cases.

\textsc{Sys} and \textsc{Red} have been implemented in the R programming language \cite{Rcore} version 4.5.3, and run on a Windows 11 Enterprise (v23H2) system equipped with an Intel Core i5-1145G7 CPU (2.60 GHz) and 32 GB of RAM. Execution times reported in Table \ref{tab:test_cases} were measured using the \texttt{microbenchmark} package \cite{microbenchmark}, with values representing the median of 100 iterations, 10 in the longer durations in boldface.

\begin{figure}[!t]
\centering
    \includegraphics[width=\linewidth, keepaspectratio]{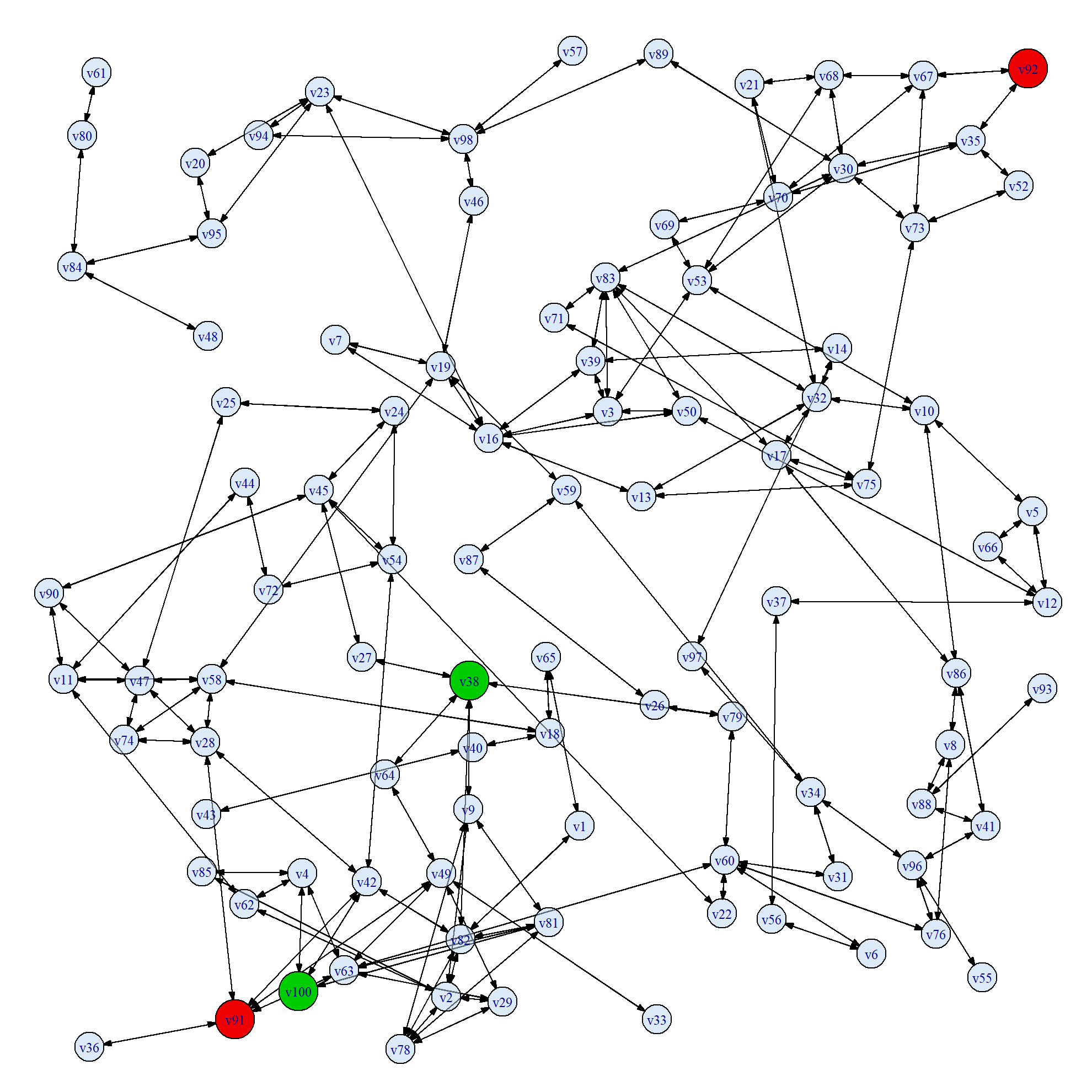}
  \caption{Network of 96 vertices, two endpoint pairs highlighted}%
  \label{fig:net96}
\end{figure}

Log-transformed values of $ProdL$ and $CutNo$ were fitted as predictors of the log execution times for \textsc{Sys} and \textsc{Red}, yielding the results presented in Table \ref{tab:linear_models}.
For a detailed guide on interpreting the regression coefficients,
$t$-statistics, and $R^2$ values presented in Table \ref{tab:linear_models}, see~\cite[chap.~7]{field2012}.

The value highlighted in underline in the \textsc{Sys} column was identified as a high-leverage outlier, attributed to system-level bottlenecks (such as garbage collection triggers) rather than algorithmic complexity, and was therefore excluded from the regression analysis.

The values in Table \ref{tab:linear_models} indicate with good confidence, although with the caveats of a small sample space, that \textsc{Sys} complexity is proportional to $ProdL^{0.46} \cdot CutNo^{0.96}$, and \textsc{Red} complexity to $ProdL^{0.23} \cdot CutNo^{0.85}$. Replacing $CutNo$ with $ProdL$ as a conservative proxy, these become $ProdL^{1.42}$ for \textsc{Sys} and $ProdL^{1.07}$ for \textsc{Red}.

This outcome of the regression is compatible with theoretical complexity calculation.

\section{Hadamard Operations}
\label{sec:hadamard}

This section introduces a vectorized formulation for the computation of pathset properties, drawing on array-programming idioms and inspired by the Iversonian philosophy~\cite{Iverson1980}. By expressing transformations through Hadamard operations and broadcasting, the formulation yields compact, language-agnostic definitions that are easy to reason about. Moreover, the vectorized form can be implemented directly on optimized numerical backends and hardware-accelerated array libraries, yielding practical performance benefits where available.

The \textit{Hadamard product}, also known as the \textit{element-wise product} or \textit{Schur product}, is a binary operation between two matrices of the same dimensions. Unlike the standard matrix multiplication, which involves summation over products of rows and columns, the Hadamard product operates entry-wise and is defined only when the operand matrices share the same size~\cite{horn2012}.

Given two matrices \( A = [a_{ij}] \) and \( B = [b_{ij}] \) of dimension \( m \times n \), their Hadamard product, denoted by \( A \odot B \), is defined as
$
(A \odot B)_{ij} = a_{ij} \cdot b_{ij}
$
, for
$
1 \leq i \leq m, \; 1 \leq j \leq n.
$

This operation is \textit{commutative}, \textit{associative}, and \textit{distributive over addition}, which make the Hadamard product particularly useful in various domains including signal processing, image analysis, machine learning, and numerical linear algebra, where component-wise operations are meaningful~\cite{marcus1964}.

Similar to the Hadamard product, the \emph{Hadamard power} of matrices $A$ and $B$ of equal dimension $m \times n$, denoted by $A^{\circ B}$, can be defined as the $m \times n$ matrix with elements given by
$
(A^{\circ B})_{ij} = a_{ij}^{b_{ij}},
$
where the elements in $A$ and $B$ are chosen such that all elements of $A^{\circ B}$ are well-defined.

\subsection{Hadamard Broadcasting}

Several scientific computing libraries, such as \texttt{numpy}, use broadcast operations. These operations allow the extension of Hadamard products and powers between matrices of the same size to matrices of different sizes. To distinguish the standard Hadamard operations for product $\odot$ and power $^\circ$ from their corresponding broadcast operations, we will, as suggested in \cite{matsui2024broadcast}, use the notations $\boxdot$ and $^\square$, respectively. This paper focuses exclusively on the Hadamard broadcast product and power operations involving a $1 \times m$ column matrix and an $m \times n$ matrix. Therefore only these specific broadcast operations will be defined.\\

\begin{Definition}
Let $A = [a_{ij}]$ be an $m \times n$ matrix, and let $\mathbf{w}$ be an $m \times 1$ column matrix over $\mathbb{R}$. The \emph{Hadamard broadcast product} of $\mathbf{w}$ and $A$, denoted by $\mathbf{w} \boxdot A$, is the $m \times n$ matrix with elements given by
$$
(\mathbf{w} \boxdot A)_{ij} = w_i \cdot a_{ij}.
$$
The \emph{Hadamard broadcast power} $\mathbf{w}$ of $A$, denoted by $\mathbf{w}^{\square A}$, is the $m \times n$ matrix with elements given by
$$
(\mathbf{w}^{\square A})_{ij} = w_i^{a_{ij}},
$$
where the elements of $\mathbf{w}$ are chosen such that all elements $w_i^{a_{ij}}$ are well-defined.
\end{Definition}

For the purposes of this paper, the convention that $0^0 = 1$~\cite{Knuth1992} will be adopted. This is consistent with the approaches of many modern programming languages, including C~\cite{IEC60559}, Java~\cite{Oracle2014}, Python~\cite{PythonSoftwareDocumentation2023}, and R~\cite{RCoreTeam2019}.

\subsection{Hadamard broadcast and r-incidence matrices}
\label{sec:equivalence}

The following proposition follows directly from the definitions of the Hadamard broadcast product and power, and of the 0-incidence and 1-incidence matrices of vertex-weighted hypergraphs.

\begin{Proposition} \label{prop:Hadamard_and_incidence_matrices}
Let $\mathcal{H}_w = (\mathcal{H}, \mathbf{w})$ be a vertex-weighted hypergraph, $B$ the $m \times n$ incidence matrix of its underlying hypergraph $\mathcal{H}$, and $\mathbf{w}$ an $m \times 1$ column vector over $\mathbb{R}$. Then,
$$
\begin{array}{cl}
(i) & \hbox{the 0-incidence matrix $B_0$ of $\mathcal{H}$ equals $\mathbf{w} \boxdot B$.}\\
(ii) & \hbox{the 1-incidence matrix $B_1$ of $\mathcal{H}$ equals $\mathbf{w}^{\square B}$.}\\
\end{array}
$$
\end{Proposition}

The Proposition~\ref{prop:Hadamard_and_incidence_matrices} allows us to establish the equivalence between Hadamard broadcast product and the r-incidence matrices.

\begin{Example}
Let $B$, $\mathbf{c}$, and $\mathbf{p}$ be defined as in examples \ref{ex:incidencematrix_cut}, \ref{ex:incidencematrix_weighted}, and \ref{ex:incidencematrix_probability_weighted} respectively.

Then, $\mathbf{c} \boxdot B$ is equal to the 0-incidence matrix $B_0 = R(0, \mathbf{c}, B)$ of $\mathcal{H}_c$, and $\mathbf{p}^{\square B}$ is equal to the 1-incidence matrix $B_1 = R(1, \mathbf{p}, B)$ of $\mathcal{H}_p$.
\end{Example}

Routing matrix, union, and cuts are incidence matrices of hypergraphs, as shown in Table~\ref{tab:transformationsummary}. Then Proposition~\ref{prop:Hadamard_and_incidence_matrices} directly implies the following theorem.

\begin{Theorem} \label{thm:transformations_incidence_matrices}
Let $P$ be a path set and $\mathbf{w}$ an edge property vector associated with the path set. If the routing table $\mathbf{R}$ equals $\routid{P}$, $\routuni{P}$, or $\routcut{P}$, then
\[
\begin{array}{ll}
   \mathbf{w} \boxdot \mathbf{R} & =  R(0, \mathbf{w}, \mathbf{R})\\
   \mathbf{w}^{\square \mathbf{R}} & = R(1, \mathbf{w}, \mathbf{R})
\end{array}
\]
\end{Theorem}

\section{Overall Implementation}
\label{sec:implementation}

Given a graph $\mathcal{G} = (V_g, E_g)$, and on it an edge property vector $\mathbf{w}$ and a path set $P$, the property of the connection implemented by the given path set can be calculated as follows.

First, an appropriate transformation of the path set is chosen and formalized as a hypergraph. The choice of the transformation is strictly related with the characteristics of the property of interest, therefore different properties will require different transformations. In section \ref{sec:previous}, three transformations with their relative hypergraphs were introduced, namely
\begin{itemize}
    \item no transformation, or $P$ as it is, $\hypid{P}$ (section \ref{sec:PathSets})
    \item union, $\hypuni{P}$ (section \ref{sec:Union})
    \item cuts, $\hypcut{P}$ (section \ref{sec:Cuts})
\end{itemize}

The transformation of choice can be represented in the form of a computer data structure as the hypergraph incidence matrix $B$. First the routing matrix $\mathbf{R}$ of $P$ is created. This is immediate by using the set definitions of the member paths of $P$. If the transformation of choice is $\hypid{P}$, this step is already completed, as $B=\routid{P}=\mathbf{R}$. In the other two cases, the incidence matrix can be calculated on $\mathbf{R}$:
\begin{itemize}
    \item if $B=\routuni{P}$, it can be obtained by taking the row-wise logical OR of zeros and ones of $\mathbf{R}$ (section \ref{sec:Union})
    \item if $B=\routcut{P}$, it can instead be calculated applying algorithm \ref{algor:cuts} to $\mathbf{R}$ (section~\ref{sec:transformations})
\end{itemize}

The next step is to augment the hypergraph into a vertex-weighted hypergraph with the edge property vector $\mathbf{w}$ and calculate its proper \(r\)-incidence matrix according with the property of interest. The value of $r$ in the \(r\)-incidence matrix shall be the identity element of the first operation to be applied, only sum and product in the considered examples, whose identity elements are 0 for the sum and 1 for the product.
The calculation of \(r\)-incidence matrices can be done taking the Hadamard product (0-incidence) or power (1-incidence) between the edge property vector $\mathbf{w}$ and the incidence matrix $B$:

\begin{equation*}
\begin{aligned}
    &B_0=\mathbf{w} \boxdot B \\
    &B_1=\mathbf{w}^{\square B}
\end{aligned}
\end{equation*}

The obtained \(r\)-incidence matrix corresponds to the matrix $\mathbf{T}$ in equation \ref{equ:MathFramework}.

Finally, the serial ($\mathit{OP}_s$) and the parallel ($\mathit{OP}_p$) composition are applied to the matrix $\mathbf{T}$ in different order, depending on the transformation.
If $\mathbf{T}=R(r, \mathbf{w}, \routid{P})$, the serial composition is applied first, column-wise, then the parallel on the resulting vector:
\begin{equation*}
    \Psi(P) = \mathit{OP}_p / (\mathit{OP}_s /) \ast \mathbf{T}
\end{equation*}
If $\mathbf{T}=R(r, \mathbf{w}, \routuni{P})$, the serial and parallel collapse in a single operator which is applied to the single column of $\mathbf{T}$:
\begin{equation*}
    \Psi(P) = \mathit{OP}_s / \mathbf{T}
\end{equation*}
If $\mathbf{T}=R(r, \mathbf{w}, \routcut{P})$, the parallel composition is applied first, column-wise, then the serial on the resulting vector:
\begin{equation*}
    \Psi(P) = \mathit{OP}_s / (\mathit{OP}_p /) \ast \mathbf{T}
\end{equation*}
where $\ast$ and $/$ represent respectively the \textit{map} and \textit{reduce} operators in Bird-Meertens formalism~\cite{Bird1987Theory}. Map applies a monadic operator to all the elements of a list, and reduce inserts a dyadic operator in between all the elements of a list. Here we follow the convention that considers matrices as lists of columns, therefore $(\mathit{OP} /) \ast$ on a matrix applies $(\mathit{OP} /)$ to all its columns, collapsing each column into the result of their elements combined with the $\mathit{OP}$ operator.

With the notations presented so far, it is now possible to provide a comprehensive summary of serial and parallel compositions, along with transformations, for calculating a property on an arbitrary set of paths. Table~\ref{tab:transformationsummary2} is the summary of mathematical model implementation for network characterization examples, allowing a direct comparison with Table~\ref{tab:transformationsummary} with the following instantiations:

\begin{table}[!t]
\centering
\resizebox{\columnwidth}{!}{%
\begin{tabular}{||l|cccc||}
 \hline
 \hline
 Example & \rule{0pt}{2.5ex} $OP_s$ & $OP_p$ & $B$ & $\mathbf{T}$ \\ [0.5ex]
 \hline
 Delay & \rule{0pt}{2.5ex} $+$ & $max$ & $\routid{P}$ & $\mathbf{w} \boxdot B$ \\
 \hline
 Administrative & \rule{0pt}{2.5ex} $+$ & $+$ & $\routuni{P}$ & $\mathbf{w} \boxdot B$ \\
 cost &  &  &  &  \\
 \hline
 Combined & \rule{0pt}{2.5ex} $min$ & $+$ & $\routcut{P}$ & $\mathbf{w} \boxdot B$ \\
 capacity &  &  &  &  \\
 \hline
 Fault & \rule{0pt}{2.5ex} $+$ & $\times$ & $\routcut{P}$ & $\mathbf{w}^{\square B}$ \\
 probability &  &  &  &  \\
 \hline
 \hline
\end{tabular}
}
\caption{Operations and transformations using Hadamard's objects}
\label{tab:transformationsummary2}
\end{table}

\begin{equation*}
    \begin{aligned}
        Delay(P) & = \max/(+/)\ast(\mathbf{w} \boxdot \routid{P})\\
         & = \max_{j}{\sum_{i}{R(0, \mathbf{w}, \routid{P})_{ij}}}\\[0.5ex]
        Cost(P) & = +/(\mathbf{w} \boxdot \routuni{P})\\
         & = \sum_{i}{(R(0, \mathbf{w}, \routuni{P})_{i}}\\[0.5ex]
        Capacity(P) & = \min/(+/)\ast(\mathbf{w} \boxdot \routcut{P})\\
         & = \min_{j}{\sum_{i}{(R(0, \mathbf{w}, \routcut{P})_{ij}}}\\[0.5ex]
        FaultProb(P) & = +/(\times /)\ast(\mathbf{w} ^{\square \routcut{P}})\\
         & = \sum_{j}{\prod_{i}{(R(1, \mathbf{w}, \routcut{P})_{ij}}}
    \end{aligned}
\end{equation*}

\section{Conclusion and Future Work}
\label{sec:Conclusion}

Building upon the formal characterization of path set properties established in~\cite{Fiaschi2025}, this paper addressed the practical implementation and computational efficiency of those definitions. The contributions are twofold:
\begin{itemize}
\item
\textbf{Algorithmic Optimization} --- The calculation of minimal cuts is non-trivial; this paper proposes two algorithms to handle this transformation efficiently.
\item
\textbf{Vectorized Computational Framework} --- By structuring property calculations as array operations, the framework adds clarity and conciseness, enabling the use of linear algebra backends.
\end{itemize}
The first contribution ensures that complex path topologies remain processable within predictable complexity bounds. The transition from the theoretical $\mathcal{O}(\mathit{ProdL})$ to an empirical performance driven by $\mathit{CutNo}$ (documented in Table~\ref{tab:linear_models}) highlights the practical efficiency of the proposed algorithms. These indicative results suggest that computing minimal cut sets remains tractable even for relatively large path sets.
Importantly, this configuration already exceeds the typical design of real-world networks, where path sets are limited to three or four paths, each spanning fewer than ten edges. The five-path case should therefore be regarded as a conservative worst case bound, well beyond what any compliant network design would require.

The second contribution, with emphasis on vectorization, aligns this work with the Iversonian tradition of array programming~\cite{Iverson1980}. By formulating path properties as matrix operations, the implementation becomes directly transferable to modern array languages, such as R, or Python/NumPy.

With the practical computability of these pathset properties, several avenues for future research emerge. The integration of these properties into active network management could be explored through:
\begin{itemize}
\item
\textbf{Heuristic Development} --- The design of multipath computation algorithms that utilize these properties directly as optimization constraints.
\item
\textbf{Multi-Objective Analysis} --- The investigation of pathset behavior when balancing several simultaneous properties, such as resilience versus latency.
\item \textbf{Dynamic Network Stability} --- The impact of time-varying network states on pathset integrity and the sensitivity of these properties to topological changes.
\end{itemize}

\section*{Acknowledgement}
The work presented in this paper is sponsored by Ericsson, Mälardalen University and the Swedish Knowledge Foundation (KKS), via the industrial PhD School ARRAY.
\balance
\bibliographystyle{IEEEtran}
\bibliography{IEEEabrv,Paperbiblio}

\end{document}